\documentclass[aps,prl,superscriptaddress,twocolumn,longbibliography,floatfix]{revtex4-2}

\usepackage{amsfonts}
\usepackage{amsmath}
\usepackage{mathrsfs}
\usepackage{amssymb}
\usepackage{graphicx}
\usepackage{float}
\usepackage{comment}
\usepackage{bm}
\usepackage{color}
\usepackage[colorlinks=true,linkcolor=blue,anchorcolor=blue,citecolor=blue,urlcolor=blue]{hyperref}
\usepackage[normalem]{ulem}
\usepackage{siunitx}
\usepackage[tracking]{microtype}
\usepackage[version=4]{mhchem}
\usepackage{elements}
\usepackage{soul}

\SetTracking{ encoding = *, shape = sc }{ 25 }

\DeclareMathSymbol{\shortminus}{\mathbin}{AMSa}{"39}

\begin{document}

\title{Observation of Momentum Space Josephson Effects}
\author{Annesh Mukhopadhyay}
\thanks{These authors contributed equally to this work}
\affiliation{Department of Physics and Astronomy, Washington State University, Pullman, WA 99164-2814, USA}
\author{Xi-Wang Luo}
\thanks{These authors contributed equally to this work}
\affiliation{CAS Key Laboratory of Quantum Information, University of Science and Technology of China, Hefei, Anhui 230026, China}
\affiliation{Synergetic Innovation Center of Quantum Information and Quantum Physics, University of Science and Technology of China, Hefei, Anhui 230026, China}
\affiliation{Hefei National Laboratory, University of Science and Technology of China, Hefei 230088, China}
\author{Colby Schimelfenig}
\affiliation{Department of Physics and Astronomy, Washington State University, Pullman, WA 99164-2814, USA}
\author{M.~K.~H. Ome}
\affiliation{Department of Physics and Astronomy, Washington State University, Pullman, WA 99164-2814, USA}
\author{Sean Mossman}
\affiliation{Department of Physics and Astronomy, Washington State University, Pullman, WA 99164-2814, USA}
\affiliation{Department of Physics and Biophysics, University of San Diego, San Diego, CA 92110, USA}
\author{Chuanwei Zhang}
\email{chuanwei.zhang@wustl.edu}
\affiliation{Department of Physics, The University of Texas at Dallas, Richardson, TX 75080-3021, USA}
\affiliation{Department of Physics, Washington University in St. Louis, St. Louis, MO 63130, USA}
\author{Peter Engels}
\email{engels@wsu.edu}
\affiliation{Department of Physics and Astronomy, Washington State University, Pullman, WA 99164-2814, USA}

\begin{abstract}
The momentum space Josephson effect describes the supercurrent flow between weakly coupled Bose-Einstein condensates (BECs) at two discrete momentum states. Here, we experimentally observe this exotic phenomenon using a BEC with Raman-induced spin-orbit coupling, where the tunneling between two local band minima is implemented by the momentum kick of an additional optical lattice. A sudden quench of the Raman detuning induces coherent spin-momentum oscillations of the BEC, which is analogous to the a.c. Josephson effect. We observe both plasma and regular Josephson oscillations in different parameter regimes. The experimental results agree well with the theoretical model and numerical simulation, and showcase the important role of nonlinear interactions. We also show that the measurement of the Josephson plasma frequency gives the Bogoliubov zero quasimomentum gap, which determines the mass of the corresponding pseudo-Goldstone mode, a long-sought phenomenon in particle physics. The observation of momentum space Josephson physics offers an exciting platform for quantum simulation and sensing utilizing momentum states as a synthetic degree.
\end{abstract}

\maketitle

\paragraph{\textcolor{blue}{Introduction.}}

The Josephson effect describes supercurrents flowing between two
reservoirs with a weak tunneling link (\textit{e.g.}, flow through a thin insulating barrier)~\cite{Josephson1962,Josephson1974}. Josephson effects have experimentally been observed in many platforms, ranging from solid state superconductors~\cite{Anderson1963} to superfluid Helium~\cite{Backhaus1997,Wheatley1975,Leggett1975,Sukhatme2001,Hoskinson2005}, exciton polaritons~\cite{Abbarchi2013}, and ultra-cold atomic gases~\cite{Albiez2005,Levy2007,Dalfovo1996,Dalfovo1999,Andrews1997,Smerzi1997,Ohberg1999,Williams1999,Raghavan1999,Cataliotti2001,Zibold2010,Kreula2017,Spagnolli2017,Burchianti2017,Burchianti2018,Valtolina2015,Luick2020}. Important applications of Josephson effects include superconducting quantum interference devices (SQUIDs)~\cite{Makhlin2001,Ryu2013}, superconducting qubits~\cite{Martinis2002,Astafiev2006,Martinis2009,Paik2011}, and precision measurements~\cite{Makhlin2001}.

In recent years, momentum states of ultracold bosons have emerged as a new synthetic degree of freedom for quantum matter and simulation. In this context, the Josephson effect in momentum space has been theoretically predicted for BECs located at two momentum states with a weak coupling induced by momentum kicks of laser beams~\cite{Huo2018}. Such momentum space tunneling has been implemented in experiments using a Bragg transition for a single component BEC~\cite{An2018,An2021} or an optical lattice in a spin-orbit coupled BEC~\cite {Bersano2019}. Despite significant experimental progress in the observation of various forms of quantum dynamics in momentum space lattices (e.g., macroscopic quantum self-trapping or phase-driven nonlinear dynamics~\cite{An2018,An2021}), the momentum space Josephson oscillation has not been observed in experiments due to the challenge of realizing a coherent ground state BEC occupying two momentum states with a long lifetime.

In this Letter, we show experimental evidence for the momentum space Josephson effect in a spin-orbit coupled BEC~\cite{Lin2011,Jim2015,Zhang2016}, whose double well band dispersion possesses two band minima at different momentum states, in analogy to real space Josephson junctions. The incorporation of a weak optical lattice induces a coupling between BECs located at two band minima, leading to the experimental observation of the long-lived ($>100$ ms) superfluid stripe ground state~\cite{Bersano2019}. Starting from the stripe ground state, a supercurrent through the momentum space junction is induced by a sudden quench of the Raman detuning between two band minima, similar to applying a voltage in a superconducting Josephson junction. The detuning quench displaces the initial stripe state from the ground state for the final detuning parameter, leading to periodic spin-momentum oscillations observed in experiments that are Josephson oscillations.

We observe two types of Josephson oscillations: \textit{i}) Josephson plasma oscillations, which are characterized by a small change in population and small phase differences between the two BECs, excited through a weak change of the system ground state; \textit{ii}) regular Josephson oscillations with a large population oscillation and a continuous increase (or decrease) of the phase difference, excited through a large change of the ground state. Our experimental results show good agreement with a theoretical model based on a two-mode approximation and numerical simulation based on the nonlinear Gross-Pitaevskii (GP) equation. The observed constant plateau of the plasma oscillation frequency in the weak lattice region showcases the important role of nonlinear interactions. Furthermore, we find that the observed Josephson plasma oscillation frequency corresponds to the zero quasimomentum gap of the Bogoliubov excitations in the superfluid stripe phase, which, in our system, represents the mass of a pseudo-Goldstone mode~\cite{Burgess2000} emerging from explicit symmetry breaking (here, the weak optical lattice breaks spatial translational symmetry). Pseudo-Goldstone modes, first proposed in particle physics~\cite{Weinberg1972}, have been a long-sought phenomenon in many different fields, and our work provides one direct experimental evidence for observing such an exotic mode.

\paragraph{\textcolor{blue}{Description of the system.}}

\begin{figure}[t]
\centering
\includegraphics[scale=1]{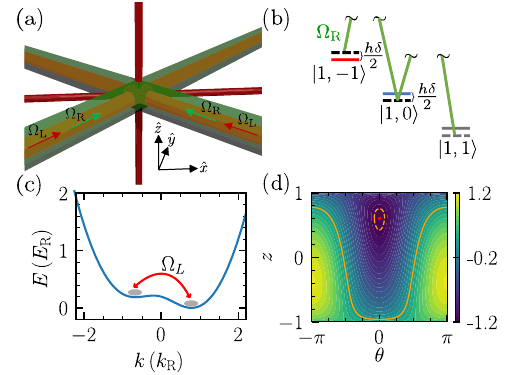}
\caption{Illustration of the experimental setup and the momentum space Josephson effect. (a) A crossed optical dipole trap (red) with two Raman laser beams (green) co-linear with two optical lattice beams (red) intersecting at the BEC position in the center. (b) Two-photon Raman transitions within the $F=1$ hyperfine manifold of $^{87}$Rb. (c) Band structure of $H_{\text{SOC}}$ for $\hbar\Omega_{\text{R}}=2.7\,E_{\text{R}}$ and $\delta=2\pi\times\qty{500}{Hz}$. (d) Phase space diagram demonstrating Josephson dynamics for $\hbar\Omega_{\text{L}}=0.5\,E_{\text{R}}$, and $gn=0.25\,E_{\text{R}}$.}
\label{fig:exp_setup}
\end{figure}

The experimental setup for the spin-orbit coupled BEC~\cite{[{See Supplementary Materials at}][{for more theoretical details about the model, Josephson \& GP dynamics, the relation with Bogoliubov spectrum, as well as some experimental details about the parameters and Bragg spectrum.}]sup_mat} has been described in our previous work~\cite{Bersano2019}. Briefly, a $^{87}$Rb BEC is confined in a cigar-shaped crossed optical dipole trap (Fig.~\ref{fig:exp_setup} (a)). An external magnetic field applied along the $x$-axis lifts the degeneracy among the three Zeeman states ($m_{F}$) in the $F=1$ hyperfine manifold. A pair of \qty{789}{nm} Raman beams intersecting at approximately $\ang{45}$ angles with the $x$-axis couples the $\lvert\uparrow\rangle\equiv|1,-1\rangle$ and $\lvert\downarrow\rangle\equiv|1,0\rangle$ Zeeman split states (Fig.~\ref{fig:exp_setup} (b)), which generates spin-orbit coupling (SOC) in the $x$-direction. The Raman coupling provides an effective momentum offset of $2\hbar k_{\text{R}}$ between these two pseudospin states. Due to the quadratic Zeeman splitting, the $|1,+1\rangle$ state is sufficiently decoupled and does not play a significant role~\cite{[{See Supplementary Materials at}][{for more theoretical details about the model, Josephson \& GP dynamics, the relation with Bogoliubov spectrum, as well as some experimental details about the parameters and Bragg spectrum.}]sup_mat}. Additionally, two \qty{1064}{nm} laser beams co-propagating with the Raman beams create a weak stationary optical lattice $V_{\text{L}}(x)=2\hbar\Omega_{\text{L}}\sin^{2}(k_{\text{L}}x)$ along the $x$-direction, which provides a $2\hbar k_{\text{L}}$ momentum kick while keeping the spin unchanged. Additional experimental details involving the atomic states and energy scales are provided in~\cite{[{See Supplementary Materials at}][{for more theoretical details about the model, Josephson \& GP dynamics, the relation with Bogoliubov spectrum, as well as some experimental details about the parameters and Bragg spectrum.}]sup_mat}.

The dynamics of the system can be described by the one-dimensional GP equation 
\begin{equation}
i\partial_{t}\psi=\left[H_{0}+g|\psi(x,t)|^{2}\right]\psi ,  \label{GP}
\end{equation}
where $\psi=\left(\psi_{\uparrow },\psi_{\downarrow}\right)^{T}$ is the two-component spinor wavefunction with the normalization to the total number of atoms $N=\int\left\vert\psi\right\vert^{2}dx$, $g$ is the density interaction strength, and $H_{0}=H_{\text{SOC}}+{V}_{\text{L}}(x)$ is the single particle Hamiltonian with
\begin{equation}
H_{\text{SOC}}=(i\partial_{x}+\sigma_{z})^{2}-\frac{\delta}{2}\sigma_{z}+\frac{\Omega_{\text{R}}}{2}\sigma_{x} .  \label{SOC}
\end{equation}
Here $\Omega_{\text{R}}$ is the Raman coupling strength, $\delta$ is the detuning of the two-photon Raman transition, and $\hbar k_{\text{R}}$ and $E_{\text{R}}=\frac{\hbar^{2}k_{\text{R}}^{2}}{2m}=h\times\qty{1.96}{kHz}$ are the momentum and energy units.

Fig.~\ref{fig:exp_setup} (c) shows the momentum-space double-well band dispersion of $H_{\text{SOC}}$. In the experiment, the period of the optical lattice is set such that $2\hbar k_{\text{L}}$ equals the separation between the two spin-orbit band minima~\cite{Bersano2019}. The optical lattice leads to the hopping between the BECs at the two band minima, in analogy to the tunneling between two superconductors separated by an insulating barrier in a Josephson junction. While the optical lattice produces multiple off-resonance couplings in addition, the momentum space Josephson junction can be more intuitively understood using a two-mode approximation, \textit{i.e.}, considering only two BEC modes at two band minima with $\psi=\left(\phi_{l}\chi_{l}e^{-ik_{\text{L}}x}+\phi_{r}\chi_{r}e^{ik_{\text{L}}x}\right)e^{ik_{\text{b}}x}$, where $\chi_{j}$ are the spinor wavefunction at two band minima, $\phi_{j}(t)$ are the mode population coefficients, and $k_{\text{b}}$ is the bias momentum induced by the detuning $\delta$. This model also neglects modes in the excited band. Denoting $\phi_{j}=\sqrt{n_{j}}e^{i\theta_{j}}$, the GP equation can be projected as
\begin{eqnarray}
\partial_{\tau}{z} &=&-\sqrt{1-z^{2}}\sin{\theta}  \notag \\
\partial_{\tau}{\theta } &=&\Lambda z+\frac{z}{\sqrt{1-z^{2}}}\cos{\theta}-\Delta E  \label{effdynamics}
\end{eqnarray}
in terms of the phase difference $\theta=\theta_{l}-\theta_{r}$ and relative population difference $z=\left(n_{r}-n_{l}\right)/n$, where $n=n_{l}+n_{r}$ is taken as a constant by neglecting populations in other modes. $\tau=2Kt$, where $K=\frac{\Omega_{\text{L}}}{2}\chi_{l}^{\ast}\chi_{r}$ describes the hopping between the two modes ($K$ is chosen to be real without loss of generality). $\Lambda=-Un/(2K)$, where $U=g|\chi_{l}^{\ast}\chi_{r}|^{2}$ represents the interaction strength of the BECs with two modes. $\Delta E=(E_{l}-E_{r})/(2K)$, with $E_{j}=\int dx\chi_{j}^{\ast}H_{0}\chi_{j}+(g+U)n$, is the energy difference between the two modes. Eq. (\ref{effdynamics}) describes a Bose Josephson junction governed by the effective Hamiltonian $H_{\text{eff}}=\frac{\Lambda}{2}z^{2}-\sqrt{1-z^{2}}\cos{\theta}-\Delta Ez$~\cite{Smerzi1997}.

A typical phase space diagram is shown in Fig.~\ref{fig:exp_setup} (d) to illustrate the momentum space Josephson dynamics. The red fixed point ($z_{0},\theta_{0}$) corresponds to the equilibrium ground state that can be obtained by finding the minima of $H_{\text{eff}}$. When the BEC is initially prepared away from ($z_{0},\theta_{0}$), ($z,\theta$) oscillate following periodic orbits in phase space, corresponding to Josephson oscillation between BECs at two band minima. There are two different types of oscillating behavior: \textit{i}) when the BEC is initially prepared not far from the fixed point, ${\theta}$ only changes within a small range for the closed periodic orbits, representing Josephson plasma oscillation shown as the orange dashed line; \textit{ii}) when the BEC is far from the fixed point, ${\theta}$ increases (or decreases) continuously through $[0,2\pi]$, corresponding to regular Josephson oscillation shown as the orange solid line~\cite{Smerzi1997}.

\paragraph{\textcolor{blue}{Observation of Josephson oscillation.}}

In our experiments, we observe momentum space Josephson dynamics after a sudden quench of the Raman detuning from an initial value $\delta_{i}$ to a final value $\delta_{f}$, in analogy to the voltage-driven ac Josephson effect. After the quench, the initially prepared superstripe state at $\delta_{i}$ is no longer the ground state at $\delta_{f}$. The BEC will then evolve under the two-mode approximation along a periodic orbit around the fixed point for $\delta_{f}$, demonstrating the Josephson oscillation. The starting point of the experiment is the preparation of a superfluid stripe state through Raman and optical lattice dressing of the BEC~\cite{Bersano2019}. SOC is generated by adiabatically ramping on the Raman beams such that the Raman coupling strength $\hbar\Omega_{\text{R}}$ increases from $0$ to $2.7\,E_{\text{R}}$ in \qty{50}{ms}. During this time, the Raman coupling is far detuned (typically, $\delta =2\pi\times(\qty{5.5}{kHz}\pm\qty{100}{Hz}$)) from the resonance. The optical lattice beams are then adiabatically applied, increasing $\hbar\Omega_{\text{L}}$ from $0$ to the desired strength in \qty{50}{ms}. Following that, $\delta$ is linearly decreased to a desired value $\delta_{i}$ in \qty{50}{ms}. This adiabatic process prepares the BEC in the superfluid stripe ground state for $\delta_{i}$.

\begin{figure}[t]
\centering
\includegraphics[scale=1]{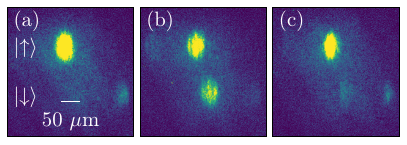}
\caption{Spin-momentum oscillation in a double-well SOC BEC. (a)--(c) Absorption images for $\hbar\Omega_{\text{L}}=0.4\,E_{\text{R}}$ after an evolution time of $t=0.15$, $1.05$, and $\qty{1.65}{ms}$.}
\label{TOFimage}
\end{figure}

In the case of a Josephson plasma oscillation, $\left(z,\theta\right)$ oscillate along a small closed orbit around $\left(z_{0},\theta_{0}\right)$ in phase space. We choose a fixed $\delta_{f}=2\pi\times\qty{500}{Hz}$ and different $\delta_{i}=2\pi\times\left(500+Q\right)$ Hz for different $\hbar\Omega_{\text{L}}\in\{0.2,0.4,\cdots,1.4\}\,E_{\text{R}}$. Proper reasoning for choosing a finite value of the final Raman detuning ($\delta_f$) can be found in~\cite{[{See Supplementary Materials at}][{for more theoretical details about the model, Josephson \& GP dynamics, the relation with Bogoliubov spectrum, as well as some experimental details about the parameters and Bragg spectrum.}]sup_mat}. Suitable values of the quench frequency $Q$ are chosen such that spin oscillations are still observable, but the initial $\left(z_{i},\theta_{i}\right)$ do not deviate significantly from $\left(z_{0},\theta_{0}\right)$, leading to plasma oscillation. After the sudden quench $\delta_{i}\rightarrow\delta_{f}$, we let the BEC evolve for a time $t$ in the presence of the Raman and optical lattice couplings. Subsequently, the Raman and lattice beams are switched off; the BEC is released from the crossed optical dipole trap, and a \qty{17.5}{ms} long time of flight (TOF), along with a briefly applied Stern-Gerlach field, resolves the BEC into different bare spin-momentum eigenstates. In the absorption images of the BEC, the two spin states are separated vertically,  and for each spin state, the momentum components are resolved horizontally~\cite{Bersano2019} (Fig.~\ref{TOFimage}). We measure the total spin polarization $\langle\sigma_{z}\rangle=(N_{\uparrow}-N_{\downarrow})/(N_{\uparrow}+N_{\downarrow})$ at each time $t$, where $N_{\uparrow}$ and $N_{\downarrow}$ are the total number of atoms in spin $\lvert\uparrow\rangle$ and $\lvert\downarrow\rangle$ respectively. Notice  that $\langle\sigma_{z}\rangle=\sum_{j}\frac{n_{j}}{n}\langle\chi_{j}|\sigma_{z}|\chi_{j}\rangle$, which can be written as $\langle\sigma_{z}\rangle=a+bz$ under the two-mode approximation, with $2a=\langle\chi_{l}|\sigma_{z}|\chi_{l}\rangle+\langle\chi_{r}|\sigma_{z}|\chi_{r}\rangle$, and $2b=\langle\chi_{r}|\sigma_{z}|\chi_{r}\rangle-\langle\chi_{l}|\sigma_{z}|\chi_{l}\rangle$. Therefore, the spin polarization oscillates with the same frequency as the Josephson oscillation. For the parameters in Fig.~\ref{fig:res_freqs}, we have $a=0.0493,b=0.732$~\cite{[{See Supplementary Materials at}][{for more theoretical details about the model, Josephson \& GP dynamics, the relation with Bogoliubov spectrum, as well as some experimental details about the parameters and Bragg spectrum.}]sup_mat}.

\begin{figure}[t]
\centering
\includegraphics[scale=1]{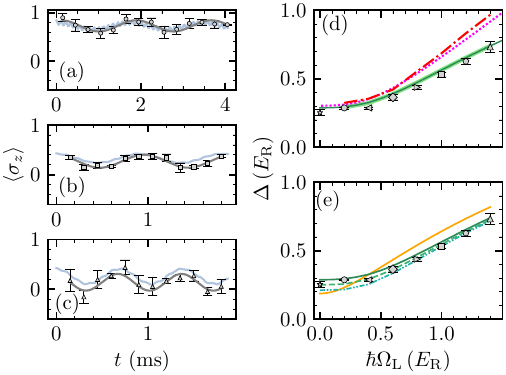}
\caption{Josephson plasma oscillation after Raman detuning quench. (a)--(c) Oscillation of the spin polarization for three different lattice coupling strengths (see main text for parameters). The solid black (light blue) curves represent the sinusoidal fitting of the experimental data (GP simulation results), while the symbols with error bars are the experimental data points. (d) Comparison between the observed $\Delta$ (symbols with error bars) and predicted $\Delta$ obtained from analyzing the Bogoliubov spectrum (solid green), quench dynamics using the GP equation (thick solid green), perturbation analysis of the two-mode Josephson model (densely dotted magenta), and the quench dynamics of Josephson model (dash-dotted red). (e) Comparison of the Bogoliubov spectrum analysis for $gn=0.25\,E_{\text{R}}$ and final Raman detuning $\delta_f=2\pi\times\qty{500}{Hz}$ (green solid line), $\qty{400}{Hz}$ (green dashed line), and $\qty{300}{Hz}$ (green densely dash-dot-dotted line) with the experimental data points for $\delta_f=2\pi\times\qty{500}{Hz}$. The solid orange line represents the calculated variation of $\Delta$ with $\hbar\Omega_{\text{L}}$ for $\delta_f=2\pi\times\qty{500}{Hz}$ and $gn=0$ (non-interacting case). The distinct markers represent the oscillation frequency experimentally obtained for the corresponding $\hbar\Omega_{\text{L}}$ from the time-dependent $\langle\sigma_{z}\rangle$ plots in (a)--(c) and similar additional plots in~\cite{[{See Supplementary Materials at}][{for more theoretical details about the model, Josephson \& GP dynamics, the relation with Bogoliubov spectrum, as well as some experimental details about the parameters and Bragg spectrum.}]sup_mat}. The star marker at $\hbar\Omega_{\text{L}}=0$ in (d) and (e) are obtained from Bragg spectroscopic measurements on a SOC BEC~\cite{[{See Supplementary Materials at}][{for more theoretical details about the model, Josephson \& GP dynamics, the relation with Bogoliubov spectrum, as well as some experimental details about the parameters and Bragg spectrum.}]sup_mat}.}
\label{fig:res_freqs}
\end{figure}

Fig.~\ref{fig:res_freqs} (a)--(c) shows the oscillations of $\langle\sigma_{z}\rangle$ during the post-quench time $t$, measured for three different lattice coupling strengths. The corresponding quench, optical lattice coupling strength, and the spin-polarization oscillation frequency $(Q,\,\hbar\Omega_{\text{L}},\,\Delta)$ for the three cases are (a) ($2\pi\times\qty{1.3}{kHz},\,0.2\,E_{\text{R}},\,(0.287\pm 0.007)\,E_{\text{R}}$), (b) ($2\pi\times\qty{200}{Hz},\,1.0\,E_{\text{R}},\,(0.530\pm 0.016)\,E_{\text{R}}$), and (c) ($2\pi\times\qty{400}{Hz},\,1.4\,E_{\text{R}},\,(0.734\pm 0.039)\,E_{\text{R}}$), where the errors in oscillation frequencies represent the standard errors in $\Delta$ obtained from the sinusoidal fitting of the corresponding data set. The experimentally observed time evolution of $\langle\sigma_{z}\rangle$ agrees reasonably well with the numerical results from directly simulating the quench dynamics using the GP Eq. (\ref{GP})~\cite{[{See Supplementary Materials at}][{for more theoretical details about the model, Josephson \& GP dynamics, the relation with Bogoliubov spectrum, as well as some experimental details about the parameters and Bragg spectrum.}]sup_mat}. In Fig. \ref{fig:res_freqs} (d), we show experimentally measured Josephson plasma oscillation frequencies with respect to $\hbar\Omega_{\text{L}}$ and their comparison with several theoretical models and numerical calculations for the interaction strength $gn=0.25\,E_{\text{R}}$. The constant Josephson plasma frequency for small values of $\hbar\Omega_{\text{L}}$ indicates a nonlinear interaction in the system, as evidenced by the comparison to the non-interacting case (Fig.~\ref{fig:res_freqs} (e)). Additionally, Fig.~\ref{fig:res_freqs} (e) shows the dependence of the plasma oscillation frequencies ($\Delta$) on the Raman detuning ($\delta$), which is an experimental parameter~\cite{[{See Supplementary Materials at}][{for more theoretical details about the model, Josephson \& GP dynamics, the relation with Bogoliubov spectrum, as well as some experimental details about the parameters and Bragg spectrum.}]sup_mat}. The different theoretical models and numerical calculations are described below:

\begin{figure}[t]
\centering
\includegraphics[scale=1]{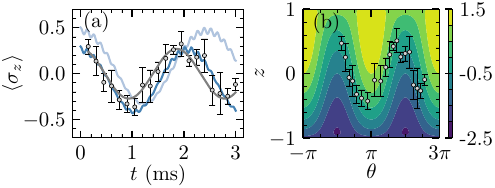}
\caption{Regular Josephson oscillation after Raman detuning quench. (a) Oscillation of the spin polarization. The solid black (light blue) curve represents a sinusoidal fit of the experimental data (GP simulation results). The dark blue curve represents the GP simulation results for a Raman detuning quench of $\delta_{i}=2\pi\times\qty{30}{Hz}$ to $\delta_{f}=-2\pi\times\qty{570}{Hz}$. (b) $z-\theta$ phase space diagram for $\hbar\Omega_{\text{L}}=0.2\,E_{\text{R}}$ and $gn=0.25\,E_{\text{R}}$ along with the experimental data. The points shown are obtained by calculating $\theta$ theoretically for the experimentally measured $z$ values according to $\langle\sigma_z\rangle=a+bz$~\cite{[{See Supplementary Materials at}][{for more theoretical details about the model, Josephson \& GP dynamics, the relation with Bogoliubov spectrum, as well as some experimental details about the parameters and Bragg spectrum.}]sup_mat}. Contour lines correspond to orbits with different energy.} 
\label{fig:josephson_osc}
\end{figure}

\textit{i}) Numerical simulations of GP Eq. (\ref{GP}), which agree well with the experimental data for all $\hbar\Omega_{\text{L}}$.

\textit{ii}) Numerical simulations of the effective dynamics (\ref{effdynamics}) from the two-mode approximation. In such detuning quench-driven dynamics, we have $\theta_{i}=0$, as seen from the fixed point position in the phase space diagram. Starting from the initial $z_{i}$ for $\delta_{i}$, we numerically integrate Eq. (\ref{effdynamics}) and determine the oscillation frequency.

\textit{iii}) Perturbation analysis of the two-mode dynamics. For the case of a small quench, we can treat the dynamics around the fixed point $(z_{0},\theta_{0}=0)$ for $\delta_{f}$ as a perturbation, \textit{i.e.}, $z=z_{0}+\delta z$, $\theta=\delta\theta$, yielding
\begin{eqnarray}
\partial_{\tau}{\delta z} &=&-\sqrt{1-z_{0}^{2}}~\delta\theta  \notag \\
\partial_{\tau}{\delta\theta} &=&\left[\Lambda+(1-z_{0}^{2})^{-3/2}\right]~\delta z,  \label{perturbation}
\end{eqnarray}
with initial conditions $\delta z(0)=z_{i}-z_{0}$ and $\delta\theta(0)=\theta_{i}-\theta_{0}=0$. Therefore the oscillation frequency $\Delta=2K\left[\sqrt{1-z_{0}^{2}}~\Lambda+\left(1-z_{0}^{2}\right)^{-1}\right]^{1/2}$. Note that the initial imbalance $z_{0}$ also depends on the lattice strength, and as $\hbar\Omega_{\text{L}}\rightarrow0$, one has $z_{0}\rightarrow-1$ and $K\rightarrow0$, leading to a finite $\Delta$. A plot of $z_0$ with respect to $\hbar\Omega_{\text{L}}$ is shown in~\cite{[{See Supplementary Materials at}][{for more theoretical details about the model, Josephson \& GP dynamics, the relation with Bogoliubov spectrum, as well as some experimental details about the parameters and Bragg spectrum.}]sup_mat}. The analytic expression agrees well with the numerical results in \textit{ii}) but deviates from the experimental results and GP simulation significantly in the large $\hbar\Omega_{\text{L}}$ regime, where higher energy modes in the SOC band are coupled by the optical lattice, leading to the failure of the two-mode approximation.

\textit{iv}) The zero quasimomentum gap of the Bogoliubov excitation spectrum. The sudden quench of the detuning leads to collective Bogoliubov quasiparticle excitations of the BEC located at quasimomentum $q=0$ due to the lack of momentum transfer. When the quench is weak, only the lowest quasiparticle band is excited. Therefore, the plasma oscillation frequency can be determined from the lowest Bogoliubov band gap. Details of this analysis and its connection with the pseudo-Goldstone mode are discussed in the next section.

While the phase varies only within a small range for a Josephson plasma oscillation, it can continuously vary through $[0,2\pi]$ for a regular Josephson oscillation. We access this regime by choosing a larger quench $\delta_{i}=2\pi\times\qty{100}{Hz}$ and $\delta_{f}=-2\pi\times\qty{500}{Hz}$ with $\hbar\Omega_{\text{L}}=0.2\,E_{\text{R}}$ so that the initial $z_{i}$ is far from the fixed point $z_{0}$. Fig.~\ref{fig:josephson_osc} (a) shows the spin-polarization oscillation as a function of the post-quench time $t$, demonstrating a large amplitude of oscillation with a change of sign in $\langle\sigma_{z}\rangle$ and an oscillation frequency of $0.275\pm 0.011\,E_{\text{R}}$. In Fig.~\ref{fig:josephson_osc} (b), we show a contour plot for the variation of the phase $\theta$ with $z$. The data points in Fig.~\ref{fig:josephson_osc} (b) are obtained by calculating $\theta$ theoretically for the experimentally measured $z$ values according to $\langle\sigma_z\rangle=a+bz$~\cite{[{See Supplementary Materials at}][{for more theoretical details about the model, Josephson \& GP dynamics, the relation with Bogoliubov spectrum, as well as some experimental details about the parameters and Bragg spectrum.}]sup_mat}. The error in $z$ represents the standard error calculated from the standard deviation in $\langle\sigma_{z}\rangle$. Fig.~\ref{fig:josephson_osc} (a) shows the regular Josephson oscillation in momentum space, with a suboptimal agreement between the theory and experiment. The dark blue curve shows the numerical GP simulation results for a Raman detuning ($\delta$) shift of $\qty{70}{Hz}$, which is within the range of $\pm\qty{100}{Hz}$ uncertainty and demonstrates a better agreement.

\paragraph{\textcolor{blue}{Connection with Bogoliubov spectrum and pseudo-Goldstone mode.}}

\begin{figure}[t]
\centering
\includegraphics[scale=1]{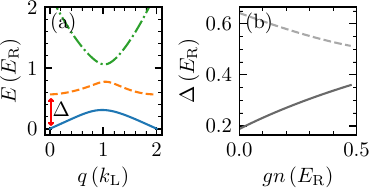}
\caption{(a) Bogoliubov spectrum of $H_{0}$ for $\hbar\Omega_{\text{R}}=2.7\,E_{\text{R}}$, $\hbar\Omega_{\text{L}}=1\,E_{\text{R}}$, $\delta=2\pi\times\qty{500}{Hz}$, and $gn=0.25\,E_{\text{R}}$. (b) Dependence of zero quasimomentum ($q=0$) band gap ($\Delta$) on $gn$ for $\hbar\Omega_{\text{L}}=0$ (solid) and $\hbar\Omega_{\text{L}}=1\,E_{\text{R}}$ (dashed).}
\label{BDG}
\end{figure}

As discussed in \emph{iv}), a small sudden Raman detuning quench generates collective excitations of the BEC from the ground band to the first excited band. The modified band structure with the incorporation of the optical lattice and mean field interaction is shown in Fig.~\ref{BDG} (a). Without the optical lattice and in the absence of a Raman detuning $\delta$, the system has translational symmetry, and the interaction leads to two gapless Goldstone modes for the stripe phase~\cite{Li2013}. The weak optical lattice breaks the translational symmetry explicitly, causing one Goldstone mode to become gapped at zero quasimomentum $q=0$. This mode is then referred to as a pseudo-Goldstone mode, which is highly relevant in the context of particle and condensed matter physics~\cite{Burgess2000}. The dependence of this pseudo-Goldstone gap on the mean-field interaction is shown in Fig.~\ref{BDG} (b). The comparison between the experimentally measured Josephson plasma oscillation frequency and the Bogoliubov excitation gap is shown in Figs.~\ref{fig:res_freqs} (d) and (e). We see that the Bogoliubov gap agrees well with the experimental measurement as well as the GP simulation. Clearly, mean-field interactions play an important role in the zero quasimomentum band gap~\cite{[{See Supplementary Materials at}][{for more theoretical details about the model, Josephson \& GP dynamics, the relation with Bogoliubov spectrum, as well as some experimental details about the parameters and Bragg spectrum.}]sup_mat}. As expected, the Bogoliubov gaps agree with those obtained from two-mode Josephson dynamics in shallow lattice regimes. The gap deviates significantly in deep lattice regimes due to the coupling with higher momentum modes.

The connection between the Josephson oscillation frequency and the pseudo-Goldstone gap can be understood as follows. Consider a small deviation of the wave function away from the ground state 
\begin{equation}
\psi=\left[(\phi_{l}^{0}+\delta\phi_{l})\chi_{l}e^{-ik_{\text{L}}x}+(\phi_{r}^{0}+\delta\phi_{r})\chi_{r}e^{ik_{\text{L}}x}\right]e^{ik_{\text{b}}x},  \label{per-BDG}
\end{equation}
that is, Bogoliubov excitations with the two-mode approximation. $\delta\phi_{l}$ and $\delta\phi_{r}$ can be obtained by solving their corresponding Bogoliubov equation~\cite{[{See Supplementary Materials at}][{for more theoretical details about the model, Josephson \& GP dynamics, the relation with Bogoliubov spectrum, as well as some experimental details about the parameters and Bragg spectrum.}]sup_mat,Li2021}. Comparing with Eq. (\ref{perturbation}), we find $\delta z=-2\sum_{j}\text{Re}[(-1)^{j}\delta\phi_{j}^{\ast}\phi_{j}^{0}]$ and $\delta\theta=-\sum_{j}\text{Im}[(-1)^{j}\delta\phi_{j}/\phi_{j}^{0}]$. By examining the two lowest modes at $q=0$, one finds that the gapped (gapless) one gives rise to non-vanishing (vanishing) $\delta z$ and $\delta\theta$. Therefore, the Josephson plasma oscillation frequency is just the zero quasimomentum Bogoliubov roton gap.

The symmetry-breaking origin of the pseudo-Goldstone mode can also be intuitively understood in the effective two-mode dynamics. Without the coupling $K$ (induced by the optical lattice), Eq.~\ref{effdynamics} has a $U_{s}(1)\times U_{a}(1)$ symmetry, where $U_{s}(1)$ corresponds to the simultaneous rotation of two modes and $U_{a}(1)$ represents equal but opposite phase rotation of two modes. The spontaneous symmetry breaking leads to two uncoupled gapless modes (\textit{i.e.}, the Goldstone modes). The introduction of a coupling $K\neq 0$ breaks the symmetry $U_{a}(1)$ and reduces the system symmetry to $U_{s}(1)$. This $U_{s}(1)$ symmetry is spontaneously broken, and the attendant Goldstone boson is absorbed and removed from the spectra via the Higgs mechanism~\cite{Esposito2007}. Only the second Goldstone boson corresponding to symmetry $U_{a}(1)$ appears. The parameter $K$ is the soft breaking parameter of the symmetry $U_{a}(1)$, and the corresponding excitations become pseudo-Goldstone Bosons.

\paragraph{\textcolor{blue}{Conclusion \& discussion.}}

Our work offers a new experimental platform for designing exotic quantum matter and engineering quantum simulators utilizing momentum states as a synthetic degree of freedom. For instance, applying Bragg scattering to the superfluid stripe ground state, one can measure the Bogoliubov excitation spectrum at finite quasimomentum, leading to the full characterization of the long-sought pseudo-Goldstone mode. Because of the spin-momentum coupling, the density interaction in the superfluid stripe phase could induce strong spin squeezing, which may be realized in our platform and used for quantum sensing. Applying small but periodic modulations of the Raman detuning can lead to the observation of a Shapiro resonance in the momentum space Josephson junction. These novel quantum phenomena enabled by the momentum space Josephson junction could potentially be useful for quantum technologies.

\begin{acknowledgments}
\textit{Acknowledgements:} A.M., C.S., M.K.H.O., S.M., and P.E. acknowledge funding from NSF through Grant No. PHY-1912540. P.E. also acknowledges support through the Ralph G. Yount Distinguished Professorship at WSU. We acknowledge experimental support from Ethan Crowell during the initial stage of this project. C.Z. is supported by AFOSR (FA9550-20-1-0220) and NSF (PHY-2409943, OMR-2228725, ECC-2411394). X.L. acknowledges support from the Innovation Program for Quantum Science and Technology (Grant No. 2021ZD0301200), the National Natural Science Foundation of China (No. 12275203), and the USTC start-up funding.
\end{acknowledgments}

\newpage

\section{Supplementary Materials}

\paragraph{\textcolor{blue}{Two-mode model and Josephson dynamics.}}

To introduce the two-mode model, we begin by writing the wavefunction of the spinor BEC in the following form
\begin{equation}
\psi(x,t)=\sum_{j}\phi_{j}(t)\chi_{j}e^{i(2j-1)k_{\text{L}}x+ik_{\text{b}}x} ,  \label{eq:wavefunction}
\end{equation}
where the integer $j=-J,...,J+1$ represent the reciprocal lattice vectors
and $J$ is the cutoff of the plane-wave modes. $\chi_{j}$ is the expansion
spinor and $\phi_{j}(t)$ are the coefficients. $k_{\text{b}}$ is the bias momentum
induced by detuning $\delta$. The dynamics are governed by 
\begin{equation}
i\partial_{t}\psi(x,t)=H_{0}\psi(x,t)+g|\psi(x,t)|^{2}\psi(x,t)  \label{eq:GP}
\end{equation}
with $g$ as the interaction strength.

\begin{figure}[h!]
    \centering
    \includegraphics[width=3.37in,height=2.05in]{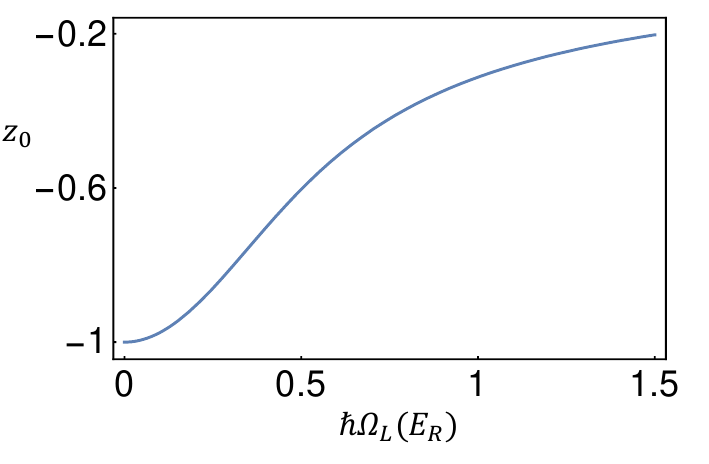}
    \caption{Initial imbalance ($z_0$) as a function of the optical lattice coupling strength ($\hbar\Omega_{\text{L}}$). All other parameters are the same as those in Fig. 3 in the main text.}
    \label{fig:imbalance_vs_lattice}
\end{figure}

We are interested in the low energy dynamics with sufficiently weak lattice and interaction strength, and thus make a two-mode approximation, that is, keeping only two modes at two band minima with $j=0,1$ and treating the lattice and interactions as perturbations. Notice that $j=0$ ($j=1$) corresponds to the left (right) minimum labeled by $l$ ($r$) in the main text. Under the two-mode approximation, we will use $j=l$ and $j=r$ to represent $j=0$ and $j=1$. The lattice couples the two modes while the interactions induce effective attractive interactions for each mode. First, we solve for the two band minima with $\Omega_{\text{L}}=0$ to obtain $k_{\text{b}}$, $k_{L}$, $\chi_{l}$, and $\chi_{r}$. Then, we rewrite the dynamic equation by substituting Eq.~\ref{eq:wavefunction} into Eq.~\ref{eq:GP} 
\begin{eqnarray}
i\partial_{t}\phi_{l}(t) &=&(E_{l}-Un_{l})\phi_{l}(t)-K\phi_{r}(t) \notag \\
i\partial_{t}\phi_{r}(t) &=&(E_{r}-Un_{r})\phi_{r}(t)-K^{\ast}\phi_{l}(t)  \label{eq:JJ}
\end{eqnarray}
where $E_{j}=\int dx\chi_{j}^{\ast}H_{0}\chi_{j}+(g+U)n$, $U=g|\chi_{l}^{\ast}\chi_{r}|^{2}$ and $K=\frac{\Omega_{\text{L}}}{2}\chi_{l}^{\ast}\chi_{r}$ (we set $K$ to be real without loss of generality), with $n=n_{l}+n_{r}$ a constant and $n_{j}=|\phi_{j}|^{2}$. The above equation is the Bose Josephson junction tunneling equation. We can write the wave function as $\phi_{j}=\sqrt{n_{j}}e^{i\theta_{j}}$, and in terms of the phase difference $\theta=\theta_{l}-\theta_{r}$ and population difference $z=\frac{n_{r}-n_{l}}{n}$, Eq.~(\ref{eq:JJ}) becomes
\begin{eqnarray}
\partial_{\tau}{z} &=&-\sqrt{1-z^{2}}\sin{\theta}  \notag \\
\partial_{\tau}{\theta} &=&\Lambda z+\frac{z}{\sqrt{1-z^{2}}}\cos{\theta}-\Delta E  \label{eq:JJ2}
\end{eqnarray}
with $\tau=2Kt$, $\Delta E=(E_{l}-E_{r})/(2K)$ and $\Lambda=-Un/(2K)$. The effective Hamiltonian reads 
\begin{equation}
H_{\text{eff}}=\frac{\Lambda}{2}z^{2}-\sqrt{1-z^{2}}\cos{\theta}-\Delta Ez.
\end{equation}
The ground state ($z_{0},\theta_{0}$) can be obtained by finding the minima of $H_{\text{eff}}$. The variation of $z_0$ with respect to $\hbar\Omega_{\text{L}}$ is shown in Fig.~\ref{fig:imbalance_vs_lattice}. Without the coupling $K$, the two-mode system Eq.~\ref{eq:JJ} has a $U_{s}(1)\times U_{a}(1)$ symmetry. The spontaneous symmetry breaking leads to two uncoupled gapless modes (i.e., the Nambu-Goldstone modes). The introduction of a coupling $K\neq 0$ breaks the symmetry $U_{a}(1)$ and reduces the system symmetry to $U_{s}(1)$ (\textit{i.e.}, simultaneous rotation of the two modes). This $U_{s}(1)$ symmetry is spontaneously broken, and the attendant Nambu-Goldstone boson is absorbed and removed from the spectra via the Higgs mechanism~\cite{Esposito2007}. Only the second Nambu-Goldstone boson corresponding to symmetry $U_{a}(1)$ (\textit{i.e.}, equal but opposite phase rotation of the two modes) appears. The parameter $K$ is the soft breaking parameter of the symmetry $U_{a}(1)$, and the corresponding excitations become pseudo Nambu-Goldstone Bosons. The frequency associated with these excitations is correspondingly small.

\paragraph{\textcolor{blue}{Treatment of quench dynamics using the GP equation.}}

Consider the spinor wave function $\psi(x,t)=[\psi_\uparrow(x,t),\psi_\downarrow(x,t)]^T$. The dynamics are characterized by the Gross-Pitaevskii (GP) equation
\begin{equation}
i\frac{\partial\psi(x,t)}{\partial t}=(H_0+V_{\text{int}})\psi(x,t) ,  \label{GPE1D}
\end{equation}
with 
\begin{equation}
V_{\text{int}}=\left(
\begin{array}{cc}
g_{\uparrow\uparrow}|\psi_\uparrow|^2+g_{\uparrow\downarrow}|\psi_\downarrow|^2 & 0 \\ 
0 & g_{\downarrow\downarrow}|\psi_\downarrow|^2+g_{\uparrow\downarrow}|\psi_\uparrow|^2
\end{array}
\right) .  \label{eq:HGP}
\end{equation}
Here, $g_{ss^{\prime}}$ denotes the interaction strength between atoms in $s$ and $s^{\prime}$ states. We can adopt the isotropic approximation for the interaction and set $g_{\uparrow\uparrow}=g_{\downarrow\downarrow}=g_{\uparrow\downarrow}=g$, which is an excellent approximation for $^{87}$Rb atoms. Before the quench, the system is in the ground state, which can be obtained by the imaginary time evolution of the GP equation. With this ground state as the initial state, the quench dynamics (\textit{i.e.}, the polarization oscillation) can be obtained by solving the real-time GP equation.

\paragraph{\textcolor{blue}{Bogoliubov spectrum.}}

To connect the Josephson oscillation frequency with the Bogoliubov spectrum, we first calculate the spectrum of Bogoliubov excitations. We write the deviations of the wavefunctions with respect to the ground state as 
\begin{equation}
\psi_{s}(x,t)=e^{-i\mu t}\left[\psi_{0s}(x)+\delta\psi_s(x,t)\right]
\label{Bogoliubov}
\end{equation}
where $\psi_{0s}(x)$ is the ground state, and the deviations take the form $\delta\psi_s(x,t)=u_{s}(x)e^{-i\varepsilon t}+v_{s}^{\ast}(x)e^{i\varepsilon t}$. The amplitudes $u_{s}(x)$ and $v_{s}(x)$ satisfy the normalization condition $\sum_{s}\int_{0}^{d}dx[|u_{s}(x)|^{2}-|v_{s}(x)|^{2}]=1$, with $d=2\pi/k_{\text{L}}$ being the lattice period and $\mu$ the chemical potential. Substituting Eq. (\ref{Bogoliubov}) into Eq. (\ref{GPE1D}), we obtain the Bogoliubov equation as $\mathcal{H}[u_{\uparrow},u_{\downarrow},v_{\uparrow},v_{\downarrow}]^T=\varepsilon[u_{\uparrow},u_{\downarrow},v_{\uparrow},v_{\downarrow}]^T$, where the Bogoliubov Hamiltonian is given by 
\begin{widetext}
\begin{equation}
\mathcal{H}=\left(
\begin{array}{cccc}
\mathcal{H}_{\uparrow} & \frac{\Omega_{\text{R}}}{2}+g_{\uparrow\downarrow}\psi_{0\uparrow}\psi_{0\downarrow}^{\ast} & g\psi_{0\uparrow}^{2} & g_{\uparrow\downarrow}\psi_{0\uparrow}\psi_{0\downarrow} \\
\frac{\Omega_{\text{R}}}{2}+g_{\uparrow\downarrow}\psi_{0\uparrow}^{\ast}\psi_{0\downarrow} & \mathcal{H}_{\downarrow} & g_{\uparrow\downarrow}\psi_{0\uparrow}\psi_{0\downarrow} & g\psi_{0\downarrow}^{2} \\
-g\psi_{0\uparrow}^{\ast 2} & -g_{\uparrow\downarrow}\psi_{0\uparrow}^{\ast}\psi_{0\downarrow}^{\ast} & -\mathcal{H}_{\uparrow}^{\ast} & -\frac{\Omega_{\text{R}}}{2}-g_{\uparrow\downarrow}\psi_{0\uparrow}^{\ast}\psi_{0\downarrow} \\
-g_{\uparrow\downarrow}\psi_{0\uparrow}^{\ast}\psi_{0\downarrow}^{\ast} & -g\psi_{0\downarrow}^{\ast 2} & -\frac{\Omega_{\text{R}}}{2}-g_{\uparrow\downarrow}\psi_{0\uparrow}\psi_{0\downarrow}^{\ast} & -\mathcal{H}_{\downarrow}^{\ast}
\end{array}
\right) ,  \label{BdGE2}
\end{equation}
with
\begin{equation}
\mathcal{H}_{\uparrow}=-\partial^{2}/\partial x^{2}+2i\partial/\partial x-\delta/2+V_{\text{L}}(x)-\mu+2g|\psi_{0\uparrow}|^{2}+g_{\uparrow\downarrow}|\psi_{0\downarrow}|^{2} ,  \label{Hup}
\end{equation}
\begin{equation}
\mathcal{H}_{\downarrow}=-\partial^{2}/\partial x^{2}-2i\partial/\partial x+\delta/2+V_{\text{L}}(x)-\mu+2g|\psi_{0\downarrow}|^{2}+g_{\uparrow\downarrow}|\psi_{0\uparrow}|^{2} .  \label{Hdown}
\end{equation}
\end{widetext}
The excitation spectra can be calculated numerically by expanding $u_{s}(x)$
and $v_{s}(x)$ in the Bloch basis. Each excitation spectrum is periodic in
momentum space with the Brillouin zone $[0,2k_{\text{L}}]$ determined by the lattice
period.

The ground state takes the following form 
\begin{equation}
\psi_0(x)=\sum_{j}\phi^0_{j}\chi_j e^{i(2j-1)k_{\text{L}}x+ik_{\text{b}}x},
\end{equation}
where the integer $j=-J,...,J+1$ represent the reciprocal lattice vectors and $J$ is the cutoff of the plane-wave modes. $\chi_j$ is the expansion spinor and $\phi^0_{j}$ are the coefficients. $k_{\text{b}}$ is the bias momentum induced by the detuning $\delta$. In the quasi-momentum frame, the two band minima become asymmetric with respect to the zero quasi-momentum if the detuning is non-zero. From the form of the ansatz, one can see that $k_b$ is located at the center of the two band minima. So when there is a non-zero detuning, $k_b$ is also nonzero~\cite{Bersano2019}.

The perturbation amplitudes $u_{s}(x)$ and $v_{s}(x)$ are expanded in the Bloch form in terms of the reciprocal lattice vectors: 
\begin{equation}
u_{s}(x)=\sum_{m=-M}^{M+1}U_{s,m}e^{i(k_\text{b}+q)x+i(2m-1)k_{\text{L}}x} ,  \label{Bogoliubov1}
\end{equation}
\begin{equation}
v_{s}(x)=\sum_{m=-M}^{M+1}V_{s,m}e^{i(k_\text{b}+q)x-i(2m-1)k_{\text{L}}x} ,  \label{Bogoliubov2}
\end{equation}
where $q$ is the Bloch wavevector of the excitations, and $M$ is the cutoff of the expansion. The whole spectrum can be obtained by substituting the expansions into the Bogoliubov equation.

The ground state of the system is a striped phase due to the presence of the optical lattice. The stripe is enforced by the optical lattice, and the translational symmetry (equivalent to the $U_a(1)$ symmetry) is explicitly broken by the lattice, which is different from the interaction-induced supersolid stripe phase where the translational symmetry is spontaneously broken. At zero-momentum, one of the two lowest Bogoliubov bands corresponds to a gapless Nambu-Goldstone mode (related to the spontaneous breaking of $U_s(1)$ symmetry), the other is a gapped pseudo-Nambu-Goldstone mode (related to the explicit breaking of $U_a(1)$ symmetry).

As mentioned above, without the lattice, the translational symmetry (\textit{i.e.}, $U_a(1)$ symmetry) may be spontaneously broken by interaction. For example, anti-ferromagnetic interaction ($g_{\uparrow\downarrow}-g_{\uparrow\uparrow}$) may lead to the supersolid stripe phase~\cite{ChunjiWang2010,TLHo2011,Lin2011,YunLi2012}; however, it is very weak, and the supersolid stripe phase only exits for extremely low values of $\Omega_{\text{R}}$ and $\delta$, making it difficult to observe in an experiment. For the dynamics studied in this paper, the effect of such anti-ferromagnetic interaction is negligible, and we can safely adopt the isotropic approach by setting $g_{\uparrow\uparrow}=g_{\downarrow\downarrow}=g_{\uparrow\downarrow}=g$.

\begin{figure}[t]
\centering
\includegraphics[width=1.0 \linewidth]{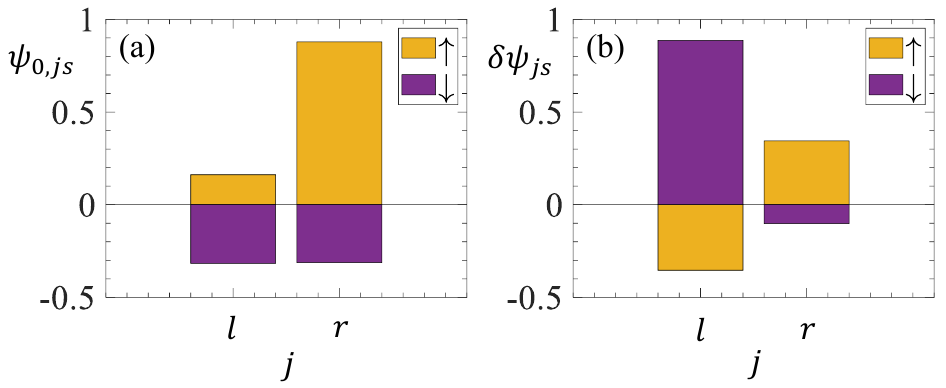}
\caption{Wave functions of the ground state and the gapped Bogoliubov excitation. (a) Wave function of ground state $\psi_{0,js}$ (unnormalized with arbitrary unit). (b) Wave function of gapped Bogoliubov excitation $\delta\psi_{js}$ (unnormalized with arbitrary unit). Parameters are $\hbar\Omega_{\text{L}}= 0.4\,E_{\text{R}}$ and $gn=0.25\,E_{\text{R}}$.}
\label{fig:BdGstate}
\end{figure}

We note that the calculation of the correct Bogoliubov spectrum requires including high momentum modes with large $J, M$. We have set $J=9$ and $M=7$ in calculating the Bogoliubov spectrum. To show that the Bogoliubov gap at $q=0$ is connected with the Josephson junction oscillation frequency, we consider only the corresponding excitation. Initially ($t=0$), the wave function of the excitation reads $\delta\psi_s(x,t=0)=u_s(x)+v^*_s(x)$, and we examine the distributions at the two band minima. That is, we consider the ground state $\psi_{0s}\simeq(\psi_{0,ls}e^{-ik_\text{L}x}+\psi_{0,rs}e^{ik_\text{L}x})e^{ik_{\text{b}}x}$ and the excitation $\delta\psi_s\simeq(\delta\psi_{ls}
e^{-ik_\text{L}x}+\delta\psi_{rs}e^{ik_\text{L}x})e^{ik_{\text{b}}x}$. The distribution amplitudes $\psi_{0,js}$ and $\delta\psi_{js}$ can be real under a proper gauge choice
and are shown in Fig.~\ref{fig:BdGstate}. We find that $\delta\psi_{ls}\propto-\psi_{0,ls}$ and $\delta\psi_{rs}\propto\psi_{0,rs}$, so we can write the  wave function as 
\begin{equation*}
\psi =\left[(\phi_{l}^{0}+\delta\phi_{l})\chi_{l}e^{-ik_{\text{L}}x}+(\phi_{r}^{0}+\delta\phi_{r})\chi_{r}e^{ik_{\text{L}}x}\right]e^{ik_{\text{b}}x} ,
\end{equation*}
and initially ($t=0$) we have $\delta\phi_l(0)\propto-\phi^0_l(0)$ and $\delta\phi_r(0)\propto\phi^0_r(0)$. We find that the gapped Bogoliubov mode corresponds to $\delta n_l\simeq-\delta n_r$ with $\delta n_j=\delta\phi_j\phi^0_j$, leading to nonzero initial $\delta z$ (notice that the initial phase difference $\delta\theta=0$ since both $\phi_j^0(0)$ and $\delta\phi_j(0)$ are real).

\paragraph{\textcolor{blue}{Explanation of oscillation frequency for $\hbar\Omega_{\text{L}}=0$.}}

The Raman detuning quench technique leads to the excitation of the gapped Bogoliubov mode (corresponding to spin excitation), and the interference with the ground state leads to spin oscillation with an identical frequency as the Bogoliubov gap. In the weak lattice region, the system is well described by the two-mode Josephson model. As one quenches the detuning, the system is quenched away from the fixed point and enters plasma oscillation (\textit{i.e.}, the spin oscillation). Therefore, in the weak lattice region, when $\hbar\Omega_{\text{L}}$ is small, both the Bogoliubov gap and the Josephson plasma frequency are determined by the observed spin oscillation. At $\hbar\Omega_{\text{L}}=0$, there would be no Josephson oscillation at all, so the numerical two-mode Josephson oscillating frequency is obtained by taking the limit $\hbar\Omega_{\text{L}}\rightarrow0$. Experimentally, this gap/frequency is obtained using Bragg spectroscopy instead of quench dynamics, where two Bragg laser beams, collinear with the Raman laser beams of varying detuning, are pulsed onto a SOC BEC. The obtained frequency from this Bragg measurement corresponds to the star marker at $\hbar\Omega_{\text{L}}=0$ in Fig. 3 (d) and (e) in the main text. Also, it is worth noting that the real-space Josephson dynamics usually rely on the tight trap and do not have an energy-momentum spectrum. Here, the momentum space Josephson dynamics does not require a real-space trap. Hence, the momentum is a good quantum number, and the pseudo-Goldstone excitation can have an energy-momentum spectrum similar to a pseudo-Goldstone boson.

\paragraph{\textcolor{blue}{Additional experimental details.}}

In our experiments, we have approximately $2.2\times 10^{5}$ atoms of $^{87}$Rb confined in a harmonic trap with trap frequencies $\bm\omega=(\omega_{x},\omega_{y},\omega_{z})=2\pi(23,154,192)$ Hz. The BEC is prepared in the $\lvert 1,-1\rangle$ Zeeman state. A \qty{10}{G} external magnetic field leads to Zeeman splitting of the $F=1$ states. The quadratic Zeeman shift for the $\lvert 1,+1\rangle$ state is \qty{14.6}{kHz}. Therefore, this state is out of resonance when near-resonant Raman coupling is applied between the $\lvert 1,-1\rangle$ and $\lvert 1,0\rangle$ states. The Raman and optical lattice coupling strengths are denoted as $\hbar\Omega_{\text{R}}$ and $\hbar\Omega_{\text{L}}$ respectively, with $\hbar k_{\text{R}}$ and $\hbar k_{\text{L}}$ being the corresponding recoil momentum where $k_{\text{L}}<k_{\text{R}}$. 

At the end of the experiment, the BEC is released from the crossed optical dipole trap and finally imaged after a 17.5 ms time of flight (TOF). The absorption images resolve the spin-momentum states, as shown in Fig. 2 in the main text. These images involve six atomic clouds, of which two are strongly populated while the others are faint due to the weak optical lattice coupling to these states. The two-photon coupling between the $\lvert 1,-1\rangle$ and $\lvert 1,0\rangle$ states effected by the Raman lasers separates the two strongly populated clouds by $2\hbar k_{\text{R}}$ in momentum space. Additionally, the optical lattice, coupling the two band minima of the SOC BEC, induces a momentum kick of $2\hbar k_{\text{L}}$ in each of the spin states, which effectively leaves the two BECs in the $\lvert\uparrow\rangle$ and $\lvert\downarrow\rangle$ spin-states separated by $2\hbar\,(k_{\text{R}}-k_{\text{L}})$ in the undressed picture. The lattice coupling to the other four states in the SOC-dressed and undressed picture is shown via schematic diagrams in~\cite{Bersano2019}.

\begin{figure}[t]
    \centering
    \includegraphics[width=0.6\linewidth]{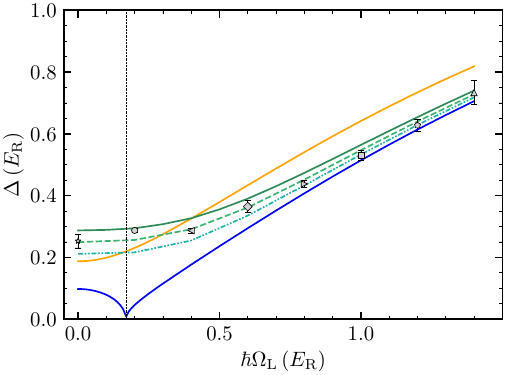}
    \caption{Comparison of the Bogoliubov spectrum analysis for $gn=0.25\,E_{\text{R}}$ and Raman detuning $\delta=2\pi\times500$ Hz (green solid line), $400$ Hz (green dashed line), and $300$ Hz (green densely dash-dot-dotted line) with the experimental data points for $\delta=2\pi\times500$ Hz. The solid orange line represents the variation of $\Delta$ with respect to $\hbar\Omega_{\text{L}}$ for $\delta=2\pi\times500$ Hz and $gn=0$ (\textit{i.e.}, single particle analysis without interactions). The blue solid line is for $\delta=0$ and $gn=0.25\,E_{\text{R}}$, where a phase transition occurs at $\hbar\Omega_{\text{L}}\simeq0.17\,E_{\text{R}}$ which is represented by the dashed vertical black line.}
    \label{fig:delta_0}
\end{figure}

\paragraph{\textcolor{blue}{Reasoning for the choice of a finite final detuning.}}
In our system, the repulsive inter-mode interaction is equivalent to attractive intra-mode interaction. Therefore, a negative curvature is expected at small $K$ values if we work at zero detuning, as shown in Fig.~\ref{fig:delta_0}. When $\delta_f=0$, the ground state would be $\left(z_0,\,\theta_0=0\right)$ at large $K$ values, where $K\propto\Omega_{\text{L}}$. Then, the plasma frequency reads $\sqrt{2K\,(2K-Un)}$, which decreases to zero at $K=Un/2$ (i.e., $\Lambda=-1$). This indicates a phase transition in the weak lattice regime shown in Fig.~\ref{fig:delta_0} by the dashed vertical black line at $\hbar\Omega_{\text{L}}\simeq0.17\,E_R$. For $K<Un/2$ (i.e., $\Lambda<-1$), the ground state has $z_0^2=1-\Lambda^{-2}\neq 0$ ($z_0$ can be negative or positive due to the spontaneous symmetry breaking). So, in the weak $K$ regime, the plasma frequency reads $\sqrt{U^2n^2-4K^2}$, which will display a negative curvature and hence a nonlinear effect. In this experiment, we work at a finite final detuning ($\delta_f=2\pi\times\qty{500}{Hz}$) due to the following reasons.

The interaction in our system is not very strong, so its effect is more significant in the weak lattice regime. However, the phase transition occurs at a weak lattice depth of $\hbar\Omega_{\text{L}}\simeq0.17\,E_R$. Therefore, in a weak lattice regime, the plasma frequency is very small in a large interval, making it hard to measure with high precision. Also, the ground state preparation is more challenging around the phase-transition point, requiring long ramp times of the parameters. Moreover, considering the $\qty{100}{Hz}$ uncertainty in our detuning, the dynamics become more sensitive to the detuning in the weak lattice regime. 

Hence, $\delta_f=0$ is not a good choice for measuring the oscillation dynamics. Also, choosing a finite final detuning does not prevent us from observing the two types of Josephson oscillation. We chose $\delta_f=2\pi\times\qty{500}{Hz}$, which is not too large but dominates over the uncertainty.

\paragraph{\textcolor{blue}{Role of atomic interactions.}} 

\begin{figure}[b]
\centering
\includegraphics[scale=1]{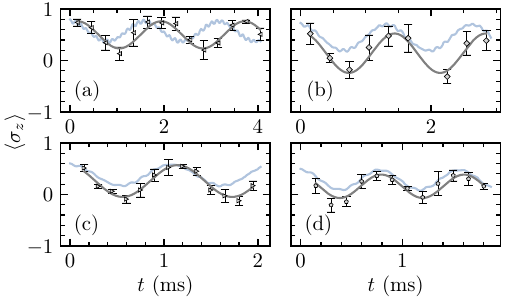}
\caption{(a)--(d) Additional data for the oscillation of $\langle\sigma_{z}\rangle$. The solid black curves represent the sinusoidal fitting of the experimental data represented by the distinct markers, and the light blue curves are the GP simulation results. See text for parameters.}
\label{fig:spin_pol}
\end{figure}

The coherent coupling of two macroscopic quantum states gives rise to the Josephson phenomenon. In this experiment, we are working with a residual detuning ($E_l\neq E_r$), and the constant plasma frequency at small $K$ (proportional to $\Omega_{\text{L}}$) values capture the interaction effect for $gn=0.25\,E_{\text{R}}$. For the non-interacting case ($gn=0$), shown by the orange line in Fig. 3 (e) in the main text, the plasma frequency would be $\sqrt{(E_l-E_r)^2+4K^2}$ in the weak lattice regime. The observed constant plateau cannot be produced from the non-interaction curve by rescaling or shifting experimental parameters.

The real-space atom-atom interaction is repulsive, but in the momentum space, it is effectively attractive. This can be seen by considering the two momentum modes. The inter-mode interaction is repulsive because when both modes are populated, the formed real-space density stripes have high interaction energy, so the BEC prefers to occupy a single momentum mode due to interaction. Effectively, this corresponds to attractive intra-mode interaction (i.e., the effective interaction is negative). As mentioned in the previous section, the interactions modify the dynamics and lead to a constant plateau of plasma frequency in a weak lattice regime, which is observed experimentally. Such a plateau is unique for the negative interaction, making it more difficult to mix the two modes. Thus, the negative interaction will weaken the effects of the coupling $K$, leading to the plateau structure.

\begin{figure}[h!]
\centering
\includegraphics[scale=1]{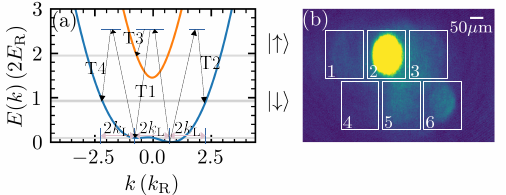}
\caption{(a) Schematic of the transitions corresponding to the observed
resonance peaks in Fig.~\ref{fig:frac_pops} (a)--(c). (b) A summed image of three absorption images corresponding to the three Bragg resonance peaks in Fig.~\ref{fig:frac_pops} (a)--(c) showing all the bare spin-momentum eigenstates $\lvert i\rangle$. Box 1, 2, 3, 5, and 6 encloses atoms in $\lvert\uparrow,\,-2\hbar k_\text{L}\,\hat{x}\rangle$, $\lvert\uparrow,\,0\rangle$, $\lvert\uparrow,\,2\hbar k_\text{L}\,\hat{x}\rangle$, $\lvert\downarrow,\,-2\hbar(k_\text{R}-k_\text{L})\,\hat{x}\rangle$, and $\lvert\downarrow,\,2\hbar k_\text{R}\,\hat{x}\rangle$ bare state, respectively. The number of atoms in box 4 is negligible.}
\label{fig:bragg_schematic}
\end{figure}

\paragraph{\textcolor{blue}{Experimental parameters.}} 
For our experimental parameters $\hbar\Omega_\text{R}=2.7\,E_{\text{R}}$, $E_{\text{R}}=1960$ Hz, $\delta=2\pi\times\qty{500}{Hz}$, and $k_{\text{L}}=\sqrt{1-(\Omega_{\text{R}}/4E_{\text{R}})^2}k_{\text{R}}$, we find the right band minimum at $k_{m_r}=0.78k_{\text{R}}$ and the left minimum at $k_{m_l}=-0.664k_{\text{R}}$, so that the ideal lattice has $k_{\text{L}}^\text{ideal}=0.722k_{\text{R}}$. However, in the realized experiment, we have used $k_{\text{L}}=\sqrt{1-(\Omega_{\text{R}}/4E_{\text{R}})^2}k_{\text{R}}=0.7378k_{\text{R}}$. Since the BEC is initially prepared with momentum $k_{m_r}$, the two modes should be at momenta $k_{m_r}=0.78k_{\text{R}}$ and $k_{m_r}-2k_{\text{L}}=-0.6956k_{\text{R}}$. Then we have $\chi_r=[0.331,-0.9436]^T$ and $\chi_l=[0.9123,-0.4096]^T$. $\Delta E=-0.1874E_{\text{R}}/(2K)$, $\Lambda=-0.4613gn/(2K)$, $2K=0.6792\Omega_{\text{L}}$. For a given $gn$ and $\Omega_{\text{L}}$, we can solve for $z_0$ and obtain the oscillation frequency. The corresponding polarization is $\langle\sigma_z\rangle=a +b z$ with coefficients $a=0.0493,b=0.732$.

We note that the dynamics are not sensitive to $k_{m_r}$ and $k_{\text{L}}$, so the effect of slight deviations of $k_{\text{L}}$ from the ideal value is negligible. On the other hand, the ideal $k_{\text{L}}$ depends on the Raman detuning. Here, we set $k_{\text{L}}$ to be the ideal value corresponding to zero detuning.

\paragraph{\textcolor{blue}{Additional plasma oscillation results.}} 
In Fig.~\ref{fig:spin_pol}, we present additional results corresponding to the Bogoliubov spectrum using the Raman detuning quench technique. This demonstrates the robust nature of our experiment, which provides access to study the dependence of $\Delta$ on $\hbar\Omega_{\text{L}}$ for a range of $\hbar\Omega_{\text{L}}$. The corresponding $Q,\,\hbar\Omega_{\text{L}},\,\Delta$ (see main text for definition) are (a) $(2\pi\times\qty{1.0}{kHz}, 0.4\,E_\text{R}, (0.287 \pm 0.008)\,E_\text{R})$, (b) $(2\pi\times\qty{900}{Hz}, 0.6\,E_\text{R}, (0.382 \pm 0.004)\,E_\text{R})$, (c) $(2\pi\times\qty{500}{Hz}, 0.8\,E_\text{R}, (0.438 \pm 0.012)\,E_\text{R})$, and (d) $(2\pi\times\qty{500}{Hz}, 1.2\,E_\text{R}, (0.628 \pm 0.02)\,E_\text{R})$. Each experimental data point presented here and in the main text is an average over three measurements, and the error bars represent the standard deviation over those measurements.

Here, we are primarily interested in short-time Josephson oscillations. Within this small time scale ($\sim$ 2 to 4 ms), decoherence induced by imperfections such as the atom loss and magnetic field fluctuation still have small effects, and coherent oscillation of the spin-polarization ($\langle\sigma_z\rangle$) is observed. At long times, the magnetic fluctuations and atom-atom scatterings would break the coherence between BECs at the two band minima, leading to damped spin oscillation for long-time dynamics.

\begin{figure}[t]
\centering
\includegraphics[scale=1]{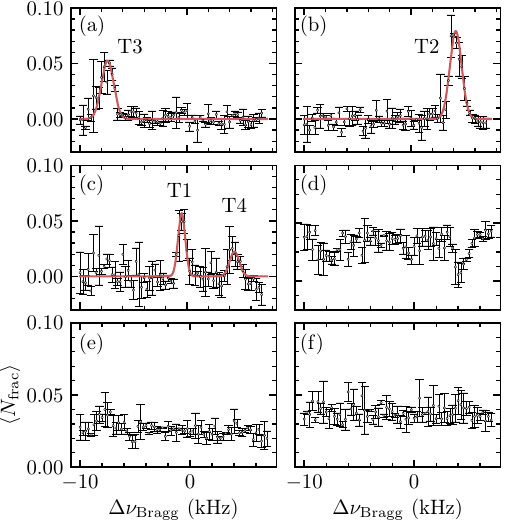}
\caption{A plot of $\langle N^\text{frac}\rangle$ with respect to $\Delta\nu_{\text{Bragg}}$ for a SOC BEC under Bragg spectroscopy. The parameters are $\hbar\Omega_\text{R}=2.7\,E_\text{R}$ and $\delta=2\pi\times\qty{500}{Hz}$. Panels (a)-(f) correspond to box no. 1, 3, 5, 2, 4, and 6 in Fig.~\ref{fig:bragg_schematic} (b), and the red lines are Gaussian fits to the peaks T3, T2, T1, and T4. Here, each gray circle is an average of over three measurements. An error bar represents the standard deviation of the three measurements. The observed resonance frequencies corresponding to the peaks are $(-7.432\pm 0.062)$ kHz, $(4.193\pm 0.031)$ kHz, $(-0.495\pm 0.044)$ kHz, and $(4.448\pm 0.141)$ kHz in (a)-(c), respectively. Panel (d) corresponds to Box 2, where most atoms were before pulsing the optical lattice. After the optical lattice pulse, atoms transfer to the other states enclosed by other boxes, and hence, there is a dip in $\langle N^\text{frac}\rangle$. Panels (e) and (f) show that the atom numbers are negligible in Box 4 and 6.}
\label{fig:frac_pops}
\end{figure}

\paragraph{\textcolor{blue}{Bragg spectroscopy.}}
\label{sec:bragg}
Experimentally, $\hbar\Omega_\text{L}=0.2\,E_\text{R}$ is the lowest lattice strength for which a clearly measurable quench
response could be obtained. To extend our measurements to zero lattice strength, Bragg spectroscopy is employed~\cite{Khamehchi2014}. For this, an optical lattice of small coupling strength ($\hbar\Omega_\text{L}=0.13\,E_\text{R}$) and of a varying detuning ($\Delta\nu_{\text{Bragg}}$) is pulsed onto the SOC BEC for $1$ ms.

The fractional population $N^\text{frac}_i$ of each bare spin-momentum eigenstate $\lvert i\rangle$ is measured after \qty{17.5}{ms} of TOF. It is defined as $N^\text{frac}_i = N_i/N_\text{total}$, where $N_i$ is the number of atoms in the eigenstate $\vert i\rangle$ and $N_\text{total}$ is the total number of atoms (Fig.~\ref{fig:bragg_schematic} (b)). Three experimental data sets are collected, and the expectation value of the fractional population $\langle N^\text{frac}_i\rangle$ is computed for each $\Delta\nu_{\text{Bragg}}$. In Fig.~\ref{fig:frac_pops}, a plot of $\langle N^\text{frac}_i\rangle$ with respect to $\Delta\nu_{\text{Bragg}}$ is shown, where each observed peak can be explained by referring to the transitions shown in Fig.~\ref{fig:bragg_schematic} (a) using the single-particle SOC band structure. The observed resonant frequencies \qty{-0.495}{kHz}, \qty{4.193}{kHz}, \qty{-7.432}{kHz}, and \qty{4.448}{kHz} correspond to the transitions T1, T2, T3, and T4 in Fig.~\ref{fig:bragg_schematic} (a), where the $(\pm)$ signs refers to positive and negative $\Delta\nu_{\text{Bragg}}$. At resonance, a positive detuning imparts a $2\hbar k_\text{L}\,\hat{x}$ momentum kick, whereas a negative detuning imparts a $-2\hbar k_\text{L}\,\hat{x}$ momentum kick. According to the single-particle SOC band structure, the predicted resonant frequencies for the transitions T1, T2, T3, and T4 are \qty{-0.368}{kHz}, \qty{3.61}{kHz}, \qty{-7.645}{kHz}, and \qty{4.448}{kHz} respectively. The difference between the theoretical values and experimental observations is attributed to the mean-field interactions. The measured energy $h\times\lvert(-0.495\pm 0.044)\rvert$ kHz associated with the transition T1 is the limiting value of the Josephson plasma frequency and the zero quasimomentum band gap for $\hbar\Omega_{\text{L}}=0$.

\end{document}